\documentclass[pre,preprint,showpacs,nonpreprintnumbers,amsmath,amssymb]{revtex4}

\usepackage{graphicx}
\usepackage{dcolumn}
\usepackage{bm}

\begin{document}

\title{Anomalous Rotational Relaxation:\\A Fractional Fokker-Planck Equation Approach}
\author{Ekrem Ayd\i ner}
\email{ekrem.aydiner@deu.edu.tr} \affiliation{ Department of
Physics, Faculty of Arts and Sciences\\Dokuz Eyl\"{u}l University,
Tr-35160 Buca, Izmir, Turkey}
\date{\today}

\begin{abstract}
In this study we have analytically obtained relaxation function in
terms of rotational correlation functions based on Brownian motion
for complex disordered systems in a stochastic framework. We found
out that rotational relaxation function has a fractional form for
complex disordered systems, which indicates relaxation has
non-exponential character obeys to Kohlrausch-William-Watts law,
following the Mittag-Leffler decay.
\end{abstract}

\maketitle Relaxation for ordered systems is given by
Maxwell-Debye law \cite{1,2} as
\begin{equation}
\Phi\left(  t\right)  =\Phi_{0}\exp\left(  -t/\tau\right)
\mathrm{,} \quad t \ge 0 \label{1}.
\end{equation}
Whereas, relaxation in many complex disordered systems such as
metallic glasses, spin glass alloys \cite{3,4,5,6}, ferroelectric
crystals \cite{7}, dielectrics \cite{8} deviates from the
classical exponential Maxwell-Debye pattern and is often described
in terms of Kohlrausch-William-Watts (KWW) (i.e. stretched
exponential) law \cite{8,9},
\begin{equation}
\Phi\left(  t\right)  =\Phi_{0}\exp\left(  -t/\tau\right)
^{\alpha} \label{2}
\end{equation}
for $0<\alpha<1$, or by an asymptotic power-law
\begin{equation}
\Phi\left(  t\right)  =\Phi_{0}\left(  1+t/\tau\right) ^{-n}
\label{3}
\end{equation}
with $n>0$. The relaxation functions in Eqs.(1) and Eq.(2) are
commonly written in terms of the correlation functions which
correspond to decay of the fluctuation of a physical quantity such
as magnetization in magnetic materials or polarization in the
dielectric materials.

Relaxation function has been derived using rotational relaxation
method for some system. This method has been firstly used by Debye
in the context of dielectric relaxation of polar molecules
\cite{2}. Debye theory is based on the Smoluchowski equation for
the non-inertial rotational diffusion of the molecules. In his
work on dielectric relaxation of an assembly of non interacting
dipolar molecules Debye considered two models of the process,
namely (a) an assembly of fixed axis rotators each having a
permanent dipole moment $\mu$ and subjected to Brownian motion
torques having their origin in the background or heat bath and (b)
the same assembly however the restriction to fixed axis rotation
is removed. The results in both instances are equivalently the
same, if inertial effects are disregarded \cite{10}. The same
picture has been applied to the rotational motion of the
magnetization vector of a super-paramagnetic particle \cite{11},
the polarization vector of a polar molecule in a dielectric
\cite{10}, and heavy molecules \cite{12,13} in the liquid and
gases. However, Debye theory cannot explain the experimental data
on dielectric relaxation of complex systems, since the
interactions between dipoles are ignored. Indeed, the relaxation
process in disordered systems is characterized by the temporal
nonlocal behavior arising from the energetic disorder which
produces obstacles or traps which delay the motion of the
particles and introduce memory effects into the motion. Therefore,
an important task in relaxation of complex systems as well as
dielectric relaxation is to extend the Debye theory of relaxation
to fractional dynamics, so that empirical decay functions, e.g.,
the stretched exponential of Kohlrausch \cite{9}, Williams and
Watts \cite{8}, may be justified. Such a generalization of the
Debye theory was given in Refs.\cite{14,15,16}. We must remark
that other generalizations of the Debye model in the context of
the fractional dynamics have been discussed in recent striking
works \cite{17,18,19,20,21} as well.

Our main aim, in this study, is obtain relaxation function Eq.(2)
for some complex disordered systems in terms of rotational
correlation functions based on rotational Brownian motion in a
stochastic framework.

The most common calculation in which the picture of rotational
Brownian motion finds relevance is that of the rotational
correlation function $\Phi_{l}^{R}\left(t\right)$, which can be
measured by Infrared (IR) and Raman (R) spectroscopies as well as
the neutron and ultrasonic scattering techniques \cite{22}.
Correlation function $\Phi_{l}^{R}\left(t\right)$ measures the
correlation in time between the direction of the unit vectors
(defining the molecular symmetry axis) $\textbf{u}(0)$ and
$\textbf{u}(t)$. Therefore, $\Phi_{l}^{R}\left(t\right)$ as can be
obtained from the IR and Raman spectroscopies may be expressed in
the compact form,
\begin{equation} 
\Phi_{l}^{R}\left(t\right)=\left\langle
P_{l}\left(\textbf{u}\left( 0\right) .\textbf{u}\left(  t\right)
\right)  \right\rangle \label{4}
\end{equation}
where $\textbf{u}$ is the unit vector along the symmetry axis of
the molecule, and $P_{l}$ is the Legendre polynomial of order $l$.
Argument of $P_{l}$ is expressed as
$\textbf{u}\left(0\right).\textbf{u}\left(t\right)=\cos\gamma\left(t\right)$,
where $\gamma$ is angle between two different point at the
spherical coordinate system. These points are given by
$\left(\theta_{0},\phi_{0}\right)$ and $\left(
 \theta,\phi\right)$ which
denote two different directions by separated by an angle $\gamma$.
There angles satisfy the trigonometric identity,
\begin{equation}
\cos\gamma=\cos\theta_{0}\cos\theta+\sin\theta_{0}\sin\theta
\cos\left(  \phi_{0}-\phi\right)\label{5}
\end{equation}
The addition theorem asserts that
\begin{equation}
P_{l}\left(\cos\gamma\right)=\frac{4\pi}{2l+1}\sum\limits_{m=-l}^{l}%
Y_{l}^{m}\left(\theta_{0},\phi_{0}\right)Y_{l}^{m\ast}\left(\theta
,\phi\right)\label{6}.
\end{equation}
If we inserted Eq.(6) into Eq.(4) rotational correlation function
is expressed in general form using spherical harmonics addition
theorem as
\begin{equation} 
\Phi_{l}\left(  t\right)  =\frac{4\pi}{2l+1}\sum\limits_{m=-l}^{l}%
\left\langle Y_{l}^{m}\left(  \theta_{0},\phi_{0}\right)
Y_{l}^{m\ast}\left( \theta,\phi\right)  \right\rangle \label{7}.
\end{equation}
Such a theoretical approach allow us to calculate for arbitrary
$l$. Average in Eq.(7) is calculated using the probability density
functions of Brownian particle which refers to end of point
polarization vector or a real particle. The Brownian motion in
disordered space in the presence of an external field
$F(x)=-V'(x)$ leads to Fractional Fokker-Planck Equation (FFPE)
\cite{23,24,25,26,27}:
\begin{equation}
\frac{\partial}{\partial t}W\left(x_{0},0\mid x,t\right)=_{0}%
D_{t}^{1-\alpha}L_{FP} W\left(x_{0},0\mid x,t\right)\label{8}
\end{equation}
\begin{equation}
L_{FP}=\left[  \frac{\partial}{\partial x}\frac{V^{\prime}\left(
x\right)}{m\eta_{\alpha}}+K_{\alpha}\frac{\partial^{2}}{\partial x^{2}%
}\right] \label{9}
\end{equation}
This equation then characterize sub-diffusion process. The FFPE
are closely related generalized L\'{e}vy-type statistics \cite{28}
and can be derived from continuous time random walk (CTRW) models
\cite{29,30,31,32,33,34}, or from a Langevin equation \cite{35}.
In Eq.(8), $W\left( x_{0},0\mid x,t\right)$ imply conditional
probability for Brownian motion, $m$ denotes mass of the particle,
$K_{\alpha}$ the diffusion constants associated with the transport
process, and the friction coefficient $\eta_{\alpha}$ is a measure
for interaction of the particle with its environment. $K_{\alpha}$
is a generalization of the Einstein-Stokes-Smoluchowski relation
\cite{23,24,25,26,27} which holds for the generalized coefficient
$\eta_{\alpha}$, which is defined as
$K_{\alpha}=k_{B}T/m\eta_{\alpha}$ where $k_{B}$ is the Boltzman
constant, and $T$ is the temperature. In the Eq.(8) operator
$_{0}D_{t}^{1-\alpha}$ is known fractional Riemann-Liouville
integro-differential operator \cite{36}.

The fractional Riemann-Liouville operator
$_{0}D_{t}^{1-\alpha}=\frac{d}{dt}_{0}D_{t}^{-\alpha}$ featuring
is defined through
\begin{eqnarray}
_{0}D_{t}^{1-\alpha}W\left(  \theta_{0},\phi_{0},0\mid\theta
,\phi,t\right)  =\frac{1}{\Gamma\left(  \alpha\right) }
\frac{\partial
}{\partial t}%
{\displaystyle\int\limits_{0}^{t}}
dt^{\prime}\frac{W\left(
\theta_{0},\phi_{0},0\mid\theta,\phi,t\right) }{\left(
t-t^{\prime}\right)^{1-\alpha}} \label{10}
\end{eqnarray}
The fractional integro-differentiation operator
$_{0}D_{t}^{1-\alpha}$ contains a convolution integral with a
slowly decaying power-law Kernel $M\left(t\right)
=\frac{t^{\alpha-1}}{\Gamma\left(  \alpha\right)}$, ensures the
non-Markovian nature of the sub-diffusion process defined by the
fractional diffusion process. Its fundamental property is the
fractional integro-differentiation of a power,
\begin{equation}
_{0}D_{t}^{1-\alpha}t^{p}=\frac{\Gamma\left(  1+p\right)  }%
{\Gamma\left(  p+\alpha\right)  }t^{p+\alpha-1} \label{11}
\end{equation}
In fact, it can be shown that more general relation
\begin{equation}
_{0}D_{t}^{p}t^{q}=\frac{\Gamma\left(  1+q\right)  }{\Gamma\left(
1+q-p\right)  }t^{q-p}  \label{12}
\end{equation}
for any real $p$, $q$. Thus, the fractional derivative of a
constant,
\begin{equation}
_{0}D_{t}^{q}1=\frac{1}{\Gamma\left(  1-q\right) }t^{-q}, \ \ \
q>0   \label{13}
\end{equation}
reproduces an inverse power-law. The special cases of integer
order integro-differentiation of a constant,
$\frac{d^{n}1}{dt^{n}}=0$, are included through the poles of the
Gamma function for $q=1,2,3,...$.

In the case of $V^{\prime}\left( x\right)=0$ which means that
there is no an external field, the one-dimensional FFPE can be
reduced to diffusive type equation;
\begin{equation}
\frac{\partial}{\partial t}W\left(  x_{0},0\mid x,t\right)  =_{0}%
D_{t}^{1-\alpha}K_{\alpha}\frac{\partial^{2}}{\partial
x^{2}}W\left( x_{0},0\mid x,t\right)\label{14}.
\end{equation}
This equation is called fractional diffusion equation
\cite{37,38,39} which is a particular form of the FFPE, which can
be represent on the spherical coordinates as a function of the
angles $\theta$ and $\phi$ as
\begin{equation}
\frac{\partial}{\partial t}W\left(
\theta_{0},\phi_{0},0\mid\theta,\phi,t\right)=_{0}D_{t}^{1-\alpha}
d_{\alpha}\nabla^{2}
W\left(\theta_{0},\phi_{0},0\mid\theta,\phi,t\right)\label{15}
\end{equation}
where $d_{\alpha}$ is referred to as the rotational diffusion
constant that is related to translational diffusion constant
$K_{\alpha}$ and the radius $a$ by $d_{\alpha}=K_{\alpha}/{a^2}$
($a=unit$)
\begin{equation}
\nabla^{2}=\frac{1}{\sin^{2}\theta}\left[  \sin\theta\frac{\partial}%
{\partial\theta}\left(
\sin\theta\frac{\partial}{\partial\theta}\right)
+\frac{\partial^{2}}{\partial\phi^{2}}\right]\label{16}
\end{equation}

The standard method of solution of Eq.(15) is the separation
variables \cite{23,24,25,26,27}. If we consider separation ansatz,
as $W=T.Q$, where $T$ and $Q$ temporal and spatial components of
the conditional probability function respectively, we will obtain
two eigen-equations as
\begin{equation}
\frac{dT\left(  0\mid
t\right)}{dt}=-\lambda^{2}d_{\alpha0}D_{t}^{1-\alpha }T\left(0\mid
t\right)\label{17}
\end{equation}
\begin{equation}
\nabla^{2}Q\left(\theta_{0},\phi_{0}\mid\theta,\phi\right)=
-\lambda^{2}Q\left(\theta_{0},\phi_{0}\mid\theta,\phi\right)\label{18}
\end{equation}
The temporal eigen-equation Eq.(17) is but the fractional
relaxation equation, the solution of which is given in terms of
the Mittag-Leffler function \cite{40}
\begin{equation}
T\left(0\mid t\right)=E_{\alpha}\left[  -d_{\alpha}l\left(
l+1\right)  t^{\alpha}\right]
\equiv\sum\limits_{j=0}^{\infty}\frac{\left[ -d_{\alpha}l\left(
l+1\right)  t^{\alpha}\right]  ^{j}}{\Gamma\left( 1+\alpha
j\right)}\label{19}.
\end{equation}
As can be seen from the series expansion, the exponential form can
be recovered in the Brownian limit $\alpha=1$
\begin{equation}
E_{1}\left[\-d_{1}l\left(\ l+1\right)\ t^{1}\right]
=\exp\left[\-d_{1}l\left(\ l+1\right)\ t\right]\label{20}.
\end{equation}
This result indicates that for $\alpha=1$ Mittag-Leffler
relaxation modes leads to Maxwell-Debye pattern, on the other
hand, for interval $0<\alpha<1$ Mittag-Leffler function indicates
non-Markovian Brownian motion.

On the other hand, a convenient way of expressing solution of the
spatial eigen-equation Eq.(18) is by means of spherical harmonics
\begin{equation}
Q\left(\theta_{0},\phi_{0}\mid\theta,\phi\right)=\sum\limits_{l=0}%
^{\infty}\sum\limits_{m=-l}^{l}\frac{4\pi}{2l+1}Y_{l}^{m}\left(
\theta _{0},\phi_{0}\right)  Y_{l}^{m\ast}\left(
\theta,\phi\right)\label{21}.
\end{equation}

As a result, if solutions Eqs.(19) and (21) are combined,
conditional probability function of Brownian particle is obtained
in terms of Mittag-Leffler function as
\begin{eqnarray}
W\left(\theta_{0},\phi_{0},0\mid\theta,\phi,t\right)
=\sum_{l=0}^{\infty}\sum_{m=-l}^{l}\frac{4\pi}{2l+1}Y_{l}^{m}\left(
\theta_{0},\phi_{0}\right) Y_{l}^{m\ast}\left(\theta,\phi\right)
E_{\alpha}\left[-d_{\alpha}l\left(\l+1\right) \
t^{\alpha}\right]\label{22}
\end{eqnarray}
Eq.(22) can be written as a compact in terms of the Euler angles
$\Omega_{0}\left(\theta_{0},\phi_{0}\right)$ and
$\Omega\left(\theta,\phi\right)$, hence
\begin{eqnarray}
W\left(  \Omega_{0},0\mid\Omega,t\right)  =\sum_{l=0}^{\infty}\sum_{m=-l}%
^{l}\frac{4\pi}{2l+1}Y_{l}^{m}\left(  \Omega_{0}\right)
Y_{l}^{m\ast}\left( \Omega\right) E_{\alpha}\left[
-d_{\alpha}l\left( \ l+1\right) \ t^{\alpha}\right]\label{23}
\end{eqnarray}
Also, Eqs.(22) and (23) should be satisfied
\begin{equation}
W\left(  \Omega_{0},0\mid\Omega,0\right)  =\delta\left(  \Omega_{0}%
-\Omega\right) \label{24}
\end{equation}
for $t=0$.

It is easy give the correlation function Eq.(7) what is by now a
familiar interpretation as
\begin{eqnarray}  
\Phi_{l}\left(  t\right)
=\frac{4\pi}{2l+1}\sum\limits_{m=-l}^{l}\int w\left(
\Omega_{0}\right)  Y_{l}^{m}\left(  \Omega_{0}\right) W\left(
\Omega_{0},0\mid\Omega,t\right) Y_{l}^{m\ast}\left( \Omega\right)
d\Omega_{0}d\Omega\label{25}.
\end{eqnarray}

In above equation $w\left(\Omega_{0}\right)$ is a priori
probability that the initial orientation is given by $\Omega_{0}$,
while $W\left(  \Omega_{0}%
,0\mid\Omega,t\right)$ is the conditional probability that the
final orientation is determined $\Omega$. Assuming that the
reorientations of the spin (or molecular) symmetry axis may be
modelled as an isotropic rotational Brownian motion, we may write
\begin{equation} 
w\left(\Omega_{0}\right)  =1/4\pi\label{26}
\end{equation}
and adopt Eq.(22) as the solution for the conditional probability.
Hence Eq.(25) yields,
\begin{eqnarray}
\Phi_{l}\left(  t\right)  =\frac{4\pi}{2l+1}\sum\limits_{m=-l}^{l}%
\sum\limits_{l^{\prime}m^{\prime}}\left(
\frac{1}{2l^{\prime}+1}\right) E_{\alpha}\left[
-d_{\alpha}l^{\prime}\left(  \ l^{\prime}+1\right) \
t^{\alpha}\right] \nonumber\\
\times \int Y\left(\Omega_{0}\right)Y_{l^{\prime}%
}^{m^{\prime}\ast}\left(\Omega_{0}\right)d\Omega_{0}\int
Y_{l}^{m}\left(\Omega\right)Y_{l^{\prime}}^{m^{\prime}\ast}\left(\Omega\right)d\Omega\label{27}
\end{eqnarray}
Using orthogonal property
\begin{equation}  
\int Y_{l}^{m}\left(  \Omega\right)
Y_{l^{\prime}}^{m^{\prime}\ast}\left( \Omega\right)
d\Omega=\delta_{ll^{\prime}}\delta_{mm^{\prime}}.\label{28}
\end{equation}
Hence, rotational correlation function (27) is written in terms of
Mittag-Leffler function \cite{40}
\begin{equation} 
\Phi_{l}\left(  t\right)  =\Phi_{l}\left(  0\right)
E_{\alpha}\left[-d_{\alpha}l\left(\l+1\right) \
t^{\alpha}\right]\label{29}
\end{equation}
where normalized factor of Eq.(29) is given as
$\Phi_{l}\left(0\right)={4\pi}/{\left(2l+1\right)^{2}}$. Eq.(29)
states simply that the rotational correlation function, starting
from the value unity at $t=0$, decay non-exponentially in time
with a relaxation time $\tau_{l}$ that inversely proportional to
rotational diffusion constant $d_{\alpha}$:
\begin{equation}  
\tau_{\alpha,l}=\left[d_{\alpha}l\left(l+1\right)\right]^{-1}\label{30}.
\end{equation}

The rotational correlation function (29) is valid for arbitrary
number $l$, and it has interesting property due to behavior of the
Mittag-Leffler functions lies in the observation that is
interpolates between an initial stretched exponential (i.e. KWW)
behavior
\begin{equation}
\Phi_{l}\left(  t\right)\approx\exp\left[-\frac{t^{\alpha}%
}{\tau_{\alpha,l}\Gamma\left(  1+\alpha\right)  }\right]\label{31}
\end{equation}
and a long-time inverse power-law pattern
\begin{equation}
\Phi_{l}\left(  t\right)  \approx\left[  \frac{\tau_{\alpha,l}t^{-\alpha}%
}{\Gamma\left(  1-\alpha\right)  }\right]. \label{32}
\end{equation}

In conclusion, it is seen that relaxation functions Eq.(31) and
(32) are compatible with Eqs.(2) and (3) for complex disordered
systems, respectively.

While drafting the problem we have disregarded the inertial of
free motion governed by the rotational kinetic energy among the
successive collision. Indeed, the rotational Brownian motion model
for complex systems make sense only when friction $\eta_{\alpha}$
is large i.e. the collisions are very rapid. There the rotational
jumps were imagined to occur by large and arbitrary angles as
opposed to the present instance in which only small-angle jumps
are considered.

In this study, we have analytically carried out the rotational
relaxation function in terms of rotational correlation function
for complex disordered systems based on rotational Brownian
motion. To obtain rotational correlation function we have
introduce generalized Fokker-Planck equations of fractional order,
which generalizes the Stokes-Einstein-Smoluchowski relation, in
consistency with the fluctuation-dissipation theorem. The
introduction of the Riemann-Liouville operator includes long-range
memory effects which are typically found in complex systems, and
consequently a single mode relaxes slowly in time, following the
Mittag-Leffler decay.

In conclude, we have shown that rotational Brownian motion in the
complex systems such as spin glasses or dielectric materials,
leads to KWW decays which indicates non-exponential relaxation.


\end{document}